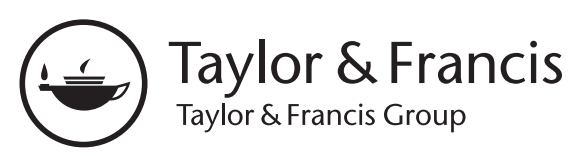

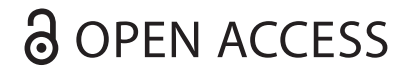

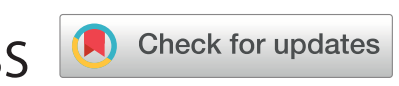

# Human-Centered Design for Connected Automation: Predicting Pedestrian Crossing Intentions

Sanaz Motamedi[a], Viktoria Marcus[b] and Griffin Pitts[c]

[a]Department of Industrial and Manufacturing Engineering, The Pennsylvania State University, University Park, PA, USA; [b]Department of Industrial and Systems Engineering, University of Florida, Gainesville, FL, USA; [c]Department of Computer and Information Science and Engineering, University of Florida, Gainesville, FL, USA

**ABSTRACT**
More than half of the 1.19 million annual traffic fatalities globally involve vulnerable road users, such as pedestrians, with a significant proportion attributable to human error. Level-5 automated driving systems (ADSs) have the potential to reduce these incidents; However, their effectiveness depends not only on automation performance but also on their ability to communicate intent and coordinate safely with pedestrians in the absence of traditional driver cues. This study aims to model pedestrian decision-making in road-crossing scenarios involving level-5 ADSs by extending the Theory of Planned Behavior (TPB) with safety, trust, compatibility, and understanding. An online survey ($n = 212$) found that perceived behavioral control, attitude, and social information significantly influence pedestrians' crossing intentions, with perceived safety and understanding exerting their influence primarily through attitude and perceived behavioral control. The results offer guidance for designing eHMIs and cooperative V2X communication strategies that promote safe pedestrian-ADS interactions and advance human-centered design for autonomous vehicles.

**KEYWORDS**
Automated driving systems; autonomous vehicles; pedestrian behavior; pedestrian–vehicle interaction; road crossing

## 1. Introduction

According to the World Health Organization (WHO), over half of the 1.19 million people who die in traffic crashes each year are vulnerable road users–those who are more likely to be severely injured or killed in a traffic crash, such as pedestrians or cyclists (WHO, 2023). Fatalities and injuries among these groups continue to rise (Naumann et al., 2025), and while the causes of traffic crashes are often difficult to determine or are caused by multiple interacting factors, studies show that a large portion are attributed to human error, involving either the driver or pedestrian (Naumann et al., 2025; WHO, 2023).

Automated vehicles have been suggested as an option to reduce human driver error by eliminating input from a human operator. According to the Society of Automotive Engineers (SAE), there are six levels of automation for automated driving systems (ADSs), levels 0–5, where level-5 ADSs are vehicles capable of "sustained and unconditional" operation in all road conditions without human supervision (Shi et al., 2020). However, level-5 ADSs will not be effective in creating safer traffic conditions for all road users if they are not well accepted by pedestrian-the most vulnerable road users. Limited research has examined the factors shaping pedestrians' willingness to interact with level-5 ADSs (Deb et al., 2017; Evans & Norman, 1998; Zhao et al., 2022).

As level-5 ADS technology becomes more mainstream, understanding pedestrians' acceptance will become increasingly important. Few studies have explored the factors impacting pedestrians' acceptance of and behavioral intention to cross the road in front of a level-5 ADS. While it is generally accepted that

**CONTACT** Sanaz Motamedi sjm7946@psu.edu Department of Industrial and Manufacturing Engineering, The Pennsylvania State University, University Park, PA, USA factors such as attitude, perceived behavioral control, trust, safety, compatibility, understanding, and social information can impact the acceptance of, and behaviors around, automated vehicles, the relationships between these factors are not well understood (Evans & Norman, 1998; Kaye et al., 2022; Marcus et al., 2024; Zhao et al., 2022, 2024). To address this gap, this study examines how safety, trust, compatibility, understanding, perceived behavioral control, attitude, and social information influence on pedestrians' behavioral intention to cross the road in front of a level-5 ADS. This study builds off the preliminary work by Marcus et al. (2024) in which a theoretical framework based on the Theory of Planned Behavior (TPB) (Ajzen, 1991) was proposed for pedestrians' acceptance and intention to cross the road in front of a level-5 ADS. In this study, the proposed theoretical framework is refined and validated using survey data.

This study contributes (1) an empirically validated extension of the TPB that specifies how safety, trust, compatibility, and understanding influence pedestrian road-crossing intentions through their effects on attitude, perceived behavioral control, and social information, and (2) evidence clarifying the relative roles and pathways of these constructs within a unified behavioral framework for level-5 automated driving interactions.

## 2. Literature review

We carried out a literature review to explore pedestrians' behavior in conventional traffic and when encountering Level-5 automated driving systems, aiming to clarify the underlying human decision-making processes.

### 2.1. Pedestrian interaction with traffic

Pedestrians are classified as vulnerable road users due to their high likelihood of injury or death in collisions (NSC, 2018), making their traffic behaviors a critical area of study. Road-crossing poses a significant risk to pedestrians and often involves complex factors, including varying environmental and infrastructural conditions (Hafeez et al., 2023; Macioszek et al., 2023; Ulfarsson et al., 2010), distractions such as mobile device use or alcohol (Ulfarsson et al., 2010), other road users and social or cultural norms (Faria et al., 2010; Pfeffer & Hunter, 2013; Rosenbloom, 2009), and factors such as age and gender (Macioszek et al., 2023; Ulfarsson et al., 2010) or personality traits such as sensation seeking, self-perception, and moral norm (Deb et al., 2017; Evans & Norman, 1998; Kaye et al., 2022; Marcus et al., 2024).

Due to the inherent hazards and complexity of traffic situations, pedestrians interpret their surroundings by reading both subtle cues and clear signals emitted by vehicles. Implicit communication, including changes in speed, distance from pedestrians, driving style, and gap size between vehicles, helps pedestrians initially understand intent (Dey et al., 2021; Pillai, 2017; Rasouli & Tsotsos, 2020; Wang et al., 2021). Next, and especially in ambiguous situations, explicit communication such as eye contact, hand waving, or other direct signals from the driver, is used to confirm intent (Pillai, 2017; Rasouli & Tsotsos, 2020; Rothenbucher et al., 2016). Pedestrians also take their cues from the actions of others, but separating the true impact of these social signals is tricky, as individuals might be responding either to one another or to the same situational cues in the environment (Faria et al., 2010; Pfeffer & Hunter, 2013; Rosenbloom, 2009; Zhao et al., 2024).

Generally, pedestrians first rely on implicit information to gauge intent, then seek explicit communication to verify it, or follow the actions of others. With a fully autonomous (Level-5) vehicle, the human driver, and the explicit cues they provide, is absent, depriving pedestrians of direct confirmation of the vehicle's intentions and potentially raising their exposure to hazard.

### 2.2. Interaction with automated driving systems (ADSs)

The absence of a human driver in level-5 ADSs raises a question about pedestrians' interactions with them, and though these vehicles are not yet widely available, initial studies offer insight. Zhao et al. (2022) reported that participants generally held favorable opinions of Level-5 autonomous vehicles and expected them to yield more readily, a belief that prompted pedestrians to engage in riskier behavior when those vehicles were present. Similarly, Jayaraman et al. (2020) found that pedestrians expected

level-5 ADSs to always stop for them, and Rothenbucher et al. (2016) found that pedestrians had high expectations of the level-5 ADS to drive safely but were more forgiving of mistakes. These attitudes, though, may lead to over-trust; if pedestrians assume the vehicle will always stop for them, they may be likely to take greater risks, such as crossing without clear yielding cues. In their observation of real-world behavior, Jayaraman et al. (2020) detected no discernible difference in how pedestrians interacted with fully autonomous Level-5 vehicles compared with human-driven ones, and in Rothenbucher et al. (2016), pedestrians who encountered a vehicle that appeared to be driverless often stepped in front of it with little hesitation, though showed more caution if the vehicle had a more aggressive driving style. Growing interest in external human–machine interfaces (eHMIs) reflects an effort to address these uncertainties by providing pedestrians with clear, accessible signals of ADS intent. eHMIs, such as text displays, symbols, speed indicators, or other visual or auditory cues, have been shown to enhance perceived safety, trust, and understanding in interactions with Level-5 ADSs by replacing the explicit communication typically provided by human drivers (Deb et al., 2018; Dey et al., 2021; Habibovic et al., 2018; Marcus et al., 2025). Although no consensus exists on optimal design, prior work suggests that well-implemented eHMIs can improve pedestrians' confidence in interpreting vehicle intent, reduce hesitation in crossing decisions, and help compensate for the loss of driver-based communication.

Beyond vehicle-related cues, the surrounding environment may also influence pedestrians' acceptance and expectations of automated vehicles. Features such as lighting conditions, weather, crosswalk design, traffic density, and roadway complexity influence how predictable and safe the setting feels, which in turn affects trust judgments and crossing decisions (MacDonald et al., 2008; Rasouli, 2020). Related work on broader infrastructure readiness similarly shows that roadway design quality and regional infrastructural conditions can influence acceptance of automated vehicles (Hafeez et al., 2024). Environments perceived as chaotic or poorly regulated may heighten uncertainty and reduce willingness to rely on autonomous vehicles, whereas well-designed, clearly signaled intersections can enhance confidence in ADS decision-making. These contextual influences point to the need for understanding not only perceptions of the vehicle itself but also how situational conditions inform pedestrians' acceptance and use of ADSs.

These studies begin exploring the differences how pedestrians interpret and respond to driverless vehicles, though questions still remain about what factors play a role in road-crossing decision making. Ongoing research into pedestrians' attitudes toward and expectations of automated vehicles will enrich our understanding of how the two are likely to interact going forward.

### 2.3. Modeling human behavior

Behavioral frameworks aim to explain how people form intentions and translate them into actions by identifying the factors that shape those intentions. Classic examples include Ajzen's Theory of Planned Behavior (1991), Davis's Technology Acceptance Model (Davis, 1989), and the Unified Theory of Acceptance and Use of Technology proposed by Venkatesh et al. (2003). Each of these models can be customized for a particular setting by adding context-specific external variables.

The TPB posits that actual behavior is largely governed by one's intention to act, and that intention is shaped by three constructs: perceived behavioral control, personal attitude toward the behavior, and perceived social (subjective) norms (Ajzen, 1991). TPB has been widely validated in the context of pedestrian interaction with automated vehicles, with PBC and attitude consistently found to be significant predictors of road-crossing intention (Evans & Norman, 1998; Kaye et al., 2022; Zhao et al., 2022). Marcus et al. (2024) proposed an extended TPB model with factors of safety, trust, compatibility, and understanding, as well as pedestrian-specific factors including general pedestrian behaviors, personal innovativeness, and background, and found that relationships between constructs, although the proposed framework was not validated. Building on Marcus et al.'s work, this study extends the TPB with external factors to examine the influences that shape pedestrians' intentions when interacting with level-5 ADSs.

### 2.4. Research questions

While previous studies have started exploring pedestrians' behavioral intention to cross the road in front of a level-5 ADS, further research is needed to clarify the roles of safety, trust, compatibility, and understanding in pedestrians' decisions, especially in risky traffic environments. This study has two main aims: (1) to examine how safety, trust, compatibility, and understanding influence the TPB model, and (2) to assess how this revised TPB model explains pedestrians' willingness to cross in front of a Level-5 autonomous vehicle.

By refining and validating the framework proposed by Marcus et al. (2024), this study examines pedestrian road-crossing decisions when interacting with level-5 ADSs, and offers insight into pedestrians' expectations and mental models, to support the creation of safer, more efficient interactions between level-5 ADSs and pedestrians.

## 3. Hypothesis development

The present work augments the TPB by adding four supplementary constructs: safety, trust, compatibility, and understanding. This framework is shown in Figure 1, and the factors and hypotheses in the model are discussed in the following sections.

### 3.1. Core TPB relationships

According to TPB, a person's desire to carry out a behavior influences their actual performance of that action, which is influenced by attitude, perceived behavioral control (PBC), and social/subjective norms (Ajzen, 1991). Because level-5 ADSs are not yet widely available, this study focuses on behavioral intention rather than actual behavior defined here as "a pedestrian's future willingness, ability, and feeling towards crossing the road in front of a level-5 ADS" (Kaye et al., 2022; Marcus et al., 2024).

A factor influencing behavioral intention in TPB is perceived behavioral control (PBC), defined as "a pedestrian's feeling of their level of control in a road-crossing scenario with a level-5 ADS, including how easy it would be to cross in front of the level-5 ADS, and if the decision were mostly up to them" (Kaye et al., 2022; Marcus et al., 2024).

Another factor impacting behavioral intention is attitude. Here attitude refers to "a pedestrian's feeling towards level-5 ADSs in general, including if they think level-5 ADSs are a good idea, would enhance the overall transportation system, and/or make roads safer for all road users" (Kaye et al., 2022; Marcus et al., 2024).

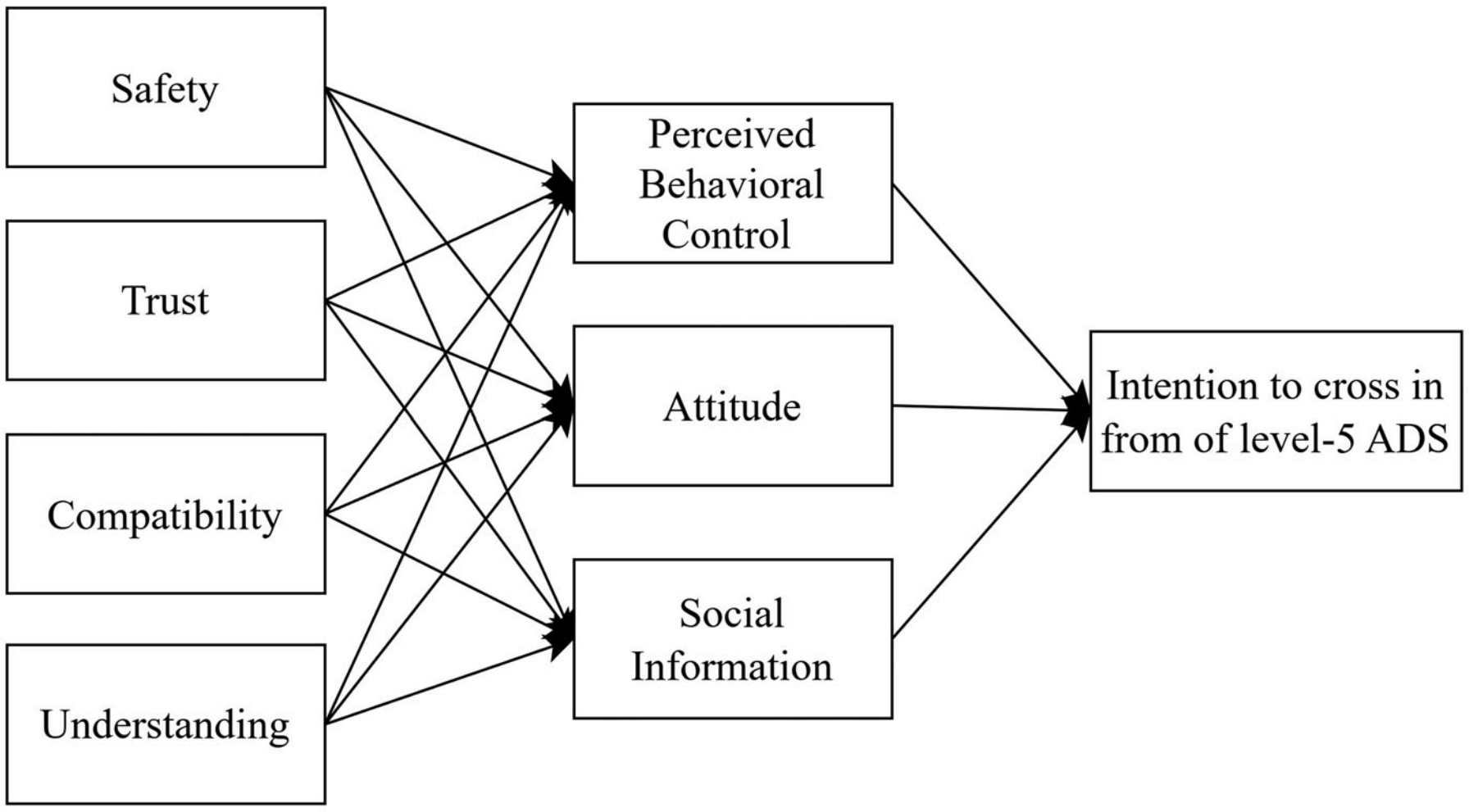

Figure 1. The theoretical framework used in this study, based on the Theory of Planned behavior and extended with additional factors of safety, trust, compatibility, and understanding, and in which social norm has been revised into social information. Arrows represent hypotheses.

The final factor included in TPB is social norms. Sometimes called subjective norms, this factor refers to "perceived social pressure to perform or not to perform [a] behavior" (Ajzen, 1991). However, Marcus et al. (2024) examined a social norm factor defined as "how strongly a pedestrian values the opinions of those important to them when deciding whether to cross the road in front of a level-5 ADS," in TPB, and found that it had no significant relationships within the TPB model. This may be due to the real-time, dynamic nature of road-crossing decisions, where pedestrians might not be as focused on broader opinions of their actions ("what will others think of me?") and instead try to gather relevant information on the traffic environment to understand the intention of the vehicle. They may also use the actions of others as an additional source of information ("what are others doing?"), as previous studies have shown that pedestrians might use social cues from other pedestrians to gather situational information and make their own road-crossing decision (Faria et al., 2010; Pfeffer & Hunter, 2013; Rosenbloom, 2009; Zhao et al., 2024). Therefore, in this study, the social norm factor is reframed as social information, defined as "the impact of other pedestrians' road-crossing decisions on one's own decision" (Faria et al., 2010; Pfeffer & Hunter, 2013; Rosenbloom, 2009; Zhao et al., 2024), shifting its role from focusing on the perception of others' opinions to their actual behaviors around the individual.

In TPB, it is expected that "the more favorable the attitude and [social] norm … and the greater the perceived behavioral control, the stronger will be an individual's intention to perform the behavior" (Ajzen, 1991). These hypotheses have been validated in the automation context by previous studies (Evans & Norman, 1998; Kaye et al., 2022; Marcus et al., 2024; Zhao et al., 2022, 2024), which leads to the following hypotheses:

- H1. Perceived behavioral control (PBC) positively affects a pedestrian's behavioral intention to cross the road in front of a level-5 ADS (BI).
- H2. Attitude (AT) positively affects a pedestrian's behavioral intention to cross the road in front of a level-5 ADS (BI).
- H3. Social information (SI) positively affects a pedestrian's behavioral intention to cross the road in front of a level-5 ADS (BI).

### 3.2. Safety

For vulnerable road users such as pedestrians, crossing the road in any traffic environment involves great personal risk. In this study, safety is defined as "how risky a pedestrian believes crossing in front of a level-5 ADS would be, their general feelings of safety around a level-5 ADS, and if they would feel more or less safe crossing the road in front of a level-5 ADS or an HDV" (Kaye et al., 2022; Marcus et al., 2024). Previous studies have shown that pedestrians' attitudes toward level-5 ADSs correspond to their feelings of safety around them, where more positive attitudes led to better perceptions of safety (Das, 2021; Rothenbucher et al., 2016; Zhao et al., 2022). Additionally, pedestrians who feel safer around level-5 ADSs generally interact with level-5 ADSs more confidently (Das, 2021; Jayaraman et al., 2020; Marcus et al., 2024; Rothenbucher et al., 2016; Zhao et al., 2022). Pedestrians may use social information as an additional means of confirming ROW or interpreting when it is safe to cross the road (Faria et al., 2010; Pfeffer & Hunter, 2013; Rosenbloom, 2009), so it is expected that when pedestrians feel more safe around a level-5 ADS, their reliance on social information will decrease. Therefore, it is hypothesized that:

- H4. Safety (S) will have a positive impact on perceived behavioral control (PBC).
- H5. Safety (S) will have a positive impact on attitude (AT).
- H6. Safety (S) will have a negative impact on social information (SI).

### 3.3. Trust

Pedestrians need confidence that a level-5 ADS detects them, interprets their intent to cross, halts accordingly, and stays stopped until they arrive at the opposite side safely. In this study, trust is defined as “the level of comfort a pedestrian would feel when considering themselves or a loved one crossing the street in front of a level-5 ADS, and their general level of trust in level-5 ADSs” (Marcus et al., 2024). In previous studies, there is consistency between higher levels of trust and positive attitudes toward level-5 ADSs, as well as pedestrians’ levels of perceived behavioral control (Jayaraman et al., 2020; Marcus et al., 2024). Similar to perceived safety, heightened trust in the approaching vehicle is expected to curb pedestrians’ reliance on social cues when deciding whether to cross the roadway (Faria et al., 2010; Pfeffer & Hunter, 2013; Rosenbloom, 2009). Therefore, it is hypothesized that:

- H7. Trust (T) will have a positive impact on perceived behavioral control (PBC).
- H8. Trust (T) will have a positive impact on attitude (AT).
- H9. Trust (T) will have a negative impact on social information (SI).

### 3.4. Compatibility

Compatibility refers to “how well a pedestrian believes that level-5 ADSs would be able to integrate into the existing traffic infrastructure and their existing road crossing habits” (Marcus et al., 2024). Previous studies have shown that pedestrians who believe level-5 ADSs are compatible with the existing traffic environment, their own crossing habits, and act like a good human driver tend to have more favorable attitudes toward them, and feel more comfortable interacting with them (Jayaraman et al., 2020; Marcus et al., 2024; Pillai, 2017; Rothenbucher et al., 2016; Zhao et al., 2022). When level-5 ADSs are compatible with existing infrastructure and pedestrians’ expectations, interactions between them become clearer, and it is expected that pedestrians’ use of social information will decrease (Faria et al., 2010; Pfeffer & Hunter, 2013; Rosenbloom, 2009). Therefore, it is hypothesized that:

- H10. Compatibility (C) will have a positive impact on perceived behavioral control (PBC).
- H11. Compatibility (C) will have a positive impact on attitude (AT).
- H12. Compatibility (C) will have a negative impact on social information (SI).

### 3.5. Understanding

Understanding is defined as “how well a pedestrian believes they would be able to interact with and understand the intentions of a level-5 ADS, and if they believe level-5 ADSs would be able to effectively communicate and interact with other pedestrians” (Marcus et al., 2024). Consistent with the effects of safety, trust, and compatibility, studies indicate that when level-5 automated vehicles convey their intentions clearly, pedestrians develop more favorable attitudes toward them and feel more at ease during interactions (Jayaraman et al., 2020; Marcus et al., 2024; Rothenbucher et al., 2016). When the intentions of a level-5 ADS are easier for pedestrians to understand, their reliance on other sources of information, including social information, is expected to decrease (Faria et al., 2010; Pfeffer & Hunter, 2013; Rosenbloom, 2009). Therefore, it is hypothesized that:

- H13. Understanding (U) will have a positive impact on perceived behavioral control (PBC).
- H14. Understanding (U) will have a positive impact on attitude (AT).
- H15. Understanding (U) will have a negative impact on social information (SI).

## 4. Methodology

### 4.1. Procedure and participants

To investigate the hypotheses in the theoretical framework, an online survey was administered using Qualtrics and structured into three sections: informed consent and demographic background, an introduction to level-5 automated driving systems (ADSs), and the theoretical framework items. The survey was distributed through research networks, social media, conference materials, and professional discussion boards. Participants were eligible to participate if they were at least 18 years old and willing to provide informed consent. No additional inclusion or exclusion criteria were applied.

Participants first read an IRB-approved consent form and then completed demographic questions, including age, gender, and disability status, following guidance from the NCWIT *Guide to Survey Demographics* (NCWIT, 2020). Respondents also answered two Likert-scale questions assessing (1) their general experience with level-5 ADSs and (2) their knowledge of the technologies used in such systems.

Next, participants viewed a subtitled two-minute educational video on level-5 ADSs, adapted from Motamedi et al. (2020) and Shi et al. (2020). The video had three primary objectives: (1) to provide insight into the various levels of automation, (2) to outline Level 5 automation and its functionalities, and (3) to present a prototype developed by Volvo Cars Corporation (Volvo Car UK, 2018). The "Next" button remained disabled until the video completed to encourage full viewing. Immediately after the video, participants completed two attention-check questions: (1) a yes/no item asking whether they had watched the full video, and (2) a content question asking them to recall the brand of the level-5 ADS shown in the introduction video. Participants were required to answer both questions correctly. If they missed either item, they were prompted to rewatch the video and then retry the same two checks. Participants who still failed either attention-check on the second attempt were automatically routed to the end of the survey. Of the 415 individuals who began the study, 15 participants (3.61%) did not pass the introduction video attention check.

Participants then responded to 33 randomized Likert-scale items measuring the constructs in the theoretical framework (four items per construct plus one repeated item for response-validation). Items were adapted from prior literature or newly developed where needed; full item wording is provided in Supplementary Appendix A (Ferenchak et al., 2022; Wu & Yuen, 2023). The repeated item was used to identify inconsistent responding, and 19 participants (4.58%) were excluded during data cleaning for selecting responses that differed by more than two (out of five) Likert points on this validation item. After all exclusions, including attention-check failures, 381 responses remained, and following completeness screening, 212 valid responses were retained for analysis.

Among the remaining 212 valid responses, participants ranged in age from 17 to 77 ($M = 28.65$, $SD = 12.22$). Gender identity was evenly split between males and females (46.70% each), with the remaining 6.60% being non-binary, self-described, or preferred not to say. Among the 12.74% of respondents reporting a disability or chronic condition, common categories included health-related (40.74%), attention-related (33.33%), and mental-health-related conditions (29.63%). On a five-point scale, respondents reported limited general experience with level-5 ADSs ($M = 1.67$, $SD = 1.15$) and slightly greater familiarity with the underlying technologies ($M = 2.16$, $SD = 1.35$).

### 4.2. Structural equation modeling

Partial least squares structural equation modeling (PLS-SEM) was conducted using SmartPLS 4 (Ringle et al., 2024). PLS-SEM was selected over covariance-based SEM because it is well suited for theory development and can accommodate smaller sample sizes, complex models, and non-normally distributed data (Hair et al., 2011). The analysis followed the recommended two-stage procedure (Hair et al., 2011; Hair, Hult, et al., 2023), beginning with assessment of the measurement model and followed by evaluation of the structural model.

For the measurement model, internal consistency reliability was assessed using both Cronbach's alpha and composite reliability (CR), with values above 0.70 considered satisfactory (Hair et al., 2019). Convergent validity was evaluated using factor loadings and average variance extracted (AVE), with AVE values above 0.50 indicating adequate convergence.

Discriminant validity was examined using the Fornell–Larcker criterion, cross-loadings, and the heterotrait–monotrait (HTMT) ratio with bootstrapped 95% confidence intervals (Fornell & Larcker, 1981; Hair, Sarstedt, et al., 2023). The Fornell–Larcker criterion evaluates whether each construct is more strongly related to its own indicators than to other constructs. This is done by comparing the square root of a construct's AVE to its correlations with all other constructs. A construct demonstrates discriminant validity when its AVE square root is higher than its correlations, indicating that it explains more variance in its own items than it shares with any other construct. In addition to the Fornell–Larcker assessment, discriminant validity was further evaluated using the heterotrait–monotrait (HTMT) ratio of correlations with bootstrapped 95% confidence intervals. HTMT provides a complementary test by estimating the degree of overlap between constructs; values below commonly recommended thresholds (0.85–0.90), along with confidence intervals that do not include 1.00, indicate adequate discriminant validity (Henseler et al., 2015). Correlations were interpreted using Cohen's (1988) benchmarks, where values of ±0.50 to ±1.00 indicate strong associations, ±0.30 to ±0.49 moderate associations, and ±0.10 to ±0.29 weak associations.

For the structural model, multicollinearity among predictor constructs was assessed using inner variance inflation factors (VIFs), with values below 5.0 indicating acceptable levels of shared variance (Hair et al., 2011; Hair, Hult, et al., 2023). Standardized path coefficients were interpreted using Keith's (2019) guidelines for small (0.05), medium (0.10), and large (0.25) effects.

Path significance was assessed using a bootstrapping procedure with 5,000 resamples, using two-tailed percentile-based 95% confidence intervals, following recommendations by Hair, Hult, et al. (2023). Model explanatory power was evaluated using $R^2$ values for each endogenous construct, interpreted according to Chin's (1998) guidelines (0.19 = weak, 0.33 = moderate, 0.67 = substantial).

Effect sizes ($f^2$) were calculated to determine the contribution of each predictor to its respective endogenous variable, with values of 0.02, 0.15, and 0.35 representing small, medium, and large effects, respectively (Cohen, 1988).

Finally, model fit was evaluated using the standardized root mean square residual (SRMR) for both the saturated and estimated models. SRMR values below 0.10 are generally considered acceptable indicators of overall model fit (Hair, Hult, et al., 2023).

## 5. Results

This section reports the PLS-SEM results following the recommended two-stage procedure (Hair, Hult, et al., 2023). The measurement model was first evaluated for reliability and validity; the structural model was then examined for multicollinearity, path significance, effect sizes, and predictive accuracy.

### 5.1. Measurement model

The measurement model was evaluated following the criteria recommended by Hair, Hult, et al. (2023), including internal consistency reliability, convergent validity, and discriminant validity. Table 1 presents the factor loadings, descriptive statistics, Cronbach's alpha coefficients, composite reliability (CR), and average variance extracted (AVE). Tables 2 and 3 present the Fornell–Larcker results and HTMT ratios.

Internal consistency reliability was supported by both Cronbach's alpha and composite reliability. Cronbach's alpha values ranged from 0.714 to 0.931, and CR values ranged from 0.794 to 0.951, all exceeding the recommended threshold of 0.70. During initial inspection, the Trust item *T2* exhibited a very low loading (0.250*), substantially reducing the construct's internal consistency ($\alpha < 0.70$). The asterisk in Table 1 identifies T2 as an item that was flagged during evaluation for having inadequate loadings. Following Hair, Hult, et al. (2023) guidance to remove indicators that substantially weaken reliability, T2 was removed and the construct was re-estimated. After removal, the Trust construct demonstrated acceptable internal consistency ($\alpha = 0.779$; CR = 0.870).

Three other items (PBC1, PBC4, and SI4) loaded below the preferred level of 0.70. These items were retained because (a) removing them did not meaningfully increase CR or AVE, and (b) the items represent theoretically important components of perceived behavioral control and social information. A sensitivity analysis confirmed that dropping these low-loading items did not change any structural path directions or significance patterns, so they were retained for conceptual completeness.

Convergent validity was assessed using factor loadings and AVE. Most items loaded strongly on their respective constructs, and all constructs achieved AVE values above 0.50 (0.503–0.828), demonstrating adequate convergent validity. After removing T2, all indicators for the Trust construct loaded acceptably and the AVE increased to 0.694.

To assess discriminant validity, we applied the Fornell–Larcker criterion (Fornell & Larcker, 1981) and examined HTMT ratios with bootstrapped confidence intervals (Hair, Sarstedt, et al., 2023). For most construct pairs, the square root of each construct's AVE was greater than its correlations with other constructs (Table 2). A small number of correlations, particularly among safety, trust, understanding, attitude, and BI, were close to or slightly higher than the lower of the two square-root AVE values. These higher correlations reflect the conceptual similarity of these perceptions. In these cases, one construct in the pair did not meet the Fornell-Larcker criterion, but the other construct did. Because the criterion is applied to each construct individually, this still indicates acceptable discriminant validity, especially given that all constructs demonstrated adequate convergent validity and theoretical distinctiveness. No constructs required removal or merging.

**Table 1.** Evaluation of survey instrument, including factor loadings, mean and standard deviations, Cronbach's alpha coefficients, CRs, and AVEs for each factor and/or item.

| Factor | Item | Factor loading | M (SD) | Cronbach's alpha | CR | AVE |
|---|---|---|---|---|---|---|
| Behavioral intention (BI) | BI1 | 0.885 | 3.33 (1.19) | 0.903 | 0.932 | 0.775 |
| | BI2 | 0.895 | 3.63 (1.23) | | | |
| | BI3 | 0.889 | 2.95 (1.23) | | | |
| | BI4 | 0.852 | 3.00 (1.18) | | | |
| Perceived behavioral control (PBC) | PBC1 | 0.588 | 3.64 (1.26) | 0.714 | 0.794 | 0.503 |
| | PBC2 | 0.841 | 3.04 (1.16) | | | |
| | PBC3 | 0.843 | 2.80 (1.30) | | | |
| | PBC4 | 0.498 | 3.56 (1.18) | | | |
| Attitude (AT) | AT1 | 0.907 | 3.37 (1.27) | 0.931 | 0.951 | 0.828 |
| | AT2 | 0.917 | 3.16 (1.22) | | | |
| | AT3 | 0.879 | 2.94 (1.32) | | | |
| | AT4 | 0.935 | 3.42 (1.26) | | | |
| Social information (SI) | SI1 | 0.816 | 3.52 (1.14) | 0.758 | 0.837 | 0.566 |
| | SI2 | 0.758 | 3.44 (1.16) | | | |
| | SI3 | 0.808 | 3.33 (1.20) | | | |
| | SI4 | 0.615 | 3.62 (1.02) | | | |
| Safety (S) | S1 | 0.900 | 3.05 (1.18) | 0.907 | 0.935 | 0.783 |
| | S2 | 0.846 | 2.85 (1.22) | | | |
| | S3 | 0.920 | 2.90 (1.22) | | | |
| | S4 | 0.871 | 2.79 (1.18) | | | |
| Trust (T) | T1 | 0.897 | 3.10 (1.27) | 0.779 | 0.870 | 0.694 |
| | T2 | 0.250* | 4.01 (1.12) | | | |
| | T3 | 0.671 | 1.89 (1.08) | | | |
| | T4 | 0.890 | 2.91 (1.21) | | | |
| Compatibility (C) | C1 | 0.804 | 3.31 (1.19) | 0.797 | 0.866 | 0.619 |
| | C2 | 0.761 | 3.06 (1.25) | | | |
| | C3 | 0.837 | 3.26 (1.23) | | | |
| | C4 | 0.741 | 2.62 (1.21) | | | |
| Understanding (U) | U1 | 0.874 | 3.17 (1.15) | 0.790 | 0.863 | 0.617 |
| | U2 | 0.785 | 2.81 (1.32) | | | |
| | U3 | 0.601 | 3.17 (1.16) | | | |
| | U4 | 0.851 | 3.15 (1.24) | | | |

**Table 2.** Inter-construct correlation matrix with square-root AVE values on the diagonal.

| Factor | BI | PBC | At | SI | S | T | C | U |
|---|---|---|---|---|---|---|---|---|
| Behavioral Intention (BI) | 0.880 | | | | | | | |
| Perceived behavioral control (PBC) | 0.728 | 0.709 | | | | | | |
| Attitude (AT) | 0.863 | 0.690 | 0.910 | | | | | |
| Social information (SI) | 0.476 | 0.359 | 0.425 | 0.752 | | | | |
| Safety (S) | 0.828 | 0.768 | 0.887 | 0.404 | 0.885 | | | |
| Trust (T) | 0.831 | 0.699 | 0.827 | 0.420 | 0.845 | 0.833 | | |
| Compatibility (C) | 0.787 | 0.728 | 0.767 | 0.437 | 0.799 | 0.730 | 0.787 | |
| Understanding (U) | 0.772 | 0.721 | 0.787 | 0.391 | 0.807 | 0.789 | 0.746 | 0.785 |

HTMT ratios provided a complementary assessment for discriminant validity. Most ratios were below the recommended 0.85–0.90 thresholds (Table 3), with elevated values occurring primarily among safety, trust, and understanding. As shown in Supplementary Appendix B, nearly all bootstrapped confidence intervals remained below 1.00; the single interval that exceeded 1.00 involved constructs that are conceptually adjacent. Together, the Fornell-Larcker and HTMT results indicate acceptable discriminant validity while acknowledging expected overlap among safety-, trust-, and understanding-related perceptions. Multicollinearity diagnostics using inner VIFs (Section 4.2) further confirmed that these relationships did not introduce problematic redundancy among constructs.

**Table 3.** HTMT ratio matrix (original sample values).

| Factor | BI | PBC | At | SI | S | T | C | U |
|---|---|---|---|---|---|---|---|---|
| Behavioral intention (BI) | — | | | | | | | |
| Perceived behavioral control (PBC) | 0.757 | — | | | | | | |
| Attitude (AT) | 0.938 | 0.738 | — | | | | | |
| Social information (SI) | 0.522 | 0.421 | 0.460 | — | | | | |
| Safety (S) | 0.905 | 0.812 | 0.953 | 0.421 | — | | | |
| Trust (T) | 0.949 | 0.732 | 0.939 | 0.472 | 0.981 | — | | |
| Compatibility (C) | 0.900 | 0.827 | 0.874 | 0.490 | 0.926 | 0.883 | — | |
| Understanding (U) | 0.900 | 0.798 | 0.899 | 0.437 | 0.935 | 0.966 | 0.900 | — |

**Table 4.** Inner model variance inflation factors (VIFs).

| Predictor | VIF |
|---|---|
| PBC -> BI | 1.930 |
| AT -> BI | 2.052 |
| SI -> BI | 1.232 |
| S -> PBC | 5.063 |
| S -> AT | 5.063 |
| S -> SI | 5.063 |
| T -> PBC | 3.982 |
| T -> AT | 3.982 |
| T -> SI | 3.982 |
| C -> PBC | 3.035 |
| C -> AT | 3.035 |
| C -> SI | 3.035 |
| U -> PBC | 3.446 |
| U -> AT | 3.446 |
| U -> SI | 3.446 |

## 5.2. Structural model

The structural model was evaluated by examining (1) the significance of the standardized path coefficients and (2) the explanatory power ($R^2$) of the endogenous constructs. Before interpreting the structural paths, multicollinearity among the predictor constructs was examined using inner variance inflation factors (VIFs). The VIFs ranged from 1.232 to 5.063. Although several values were close to the commonly cited upper guideline of 5.0 (Hair et al., 2011; Hair, Hult, et al., 2023), they did not indicate substantive multicollinearity concerns. The highest VIF values were associated with predictors that are theoretically related, which can naturally produce stronger shared variance. Overall, the VIF results support the stability and interpretability of the estimated structural relationships (Table 4).

Proceeding with the PLS-SEM results, BI was significantly predicted by PBC ($\beta = 0.242$, $p < 0.001$), attitude ($\beta = 0.647$, $p < 0.001$), and SI ($\beta = 0.115$, $p = 0.006$). Among the antecedents of these factors, safety exerted significant positive effects on attitude ($\beta = 0.542$, $p < 0.001$) and PBC ($\beta = 0.359$, $p = 0.001$), but not on SI. Trust significantly affected attitude ($\beta = 0.211$, $p = 0.001$), but its effects on PBC and SI were not significant. Understanding significantly predicted both attitude ($\beta = 0.110$, $p = 0.047$) and PBC ($\beta = 0.203$, $p = 0.022$), but not SI. Compatibility significantly predicted PBC ($\beta = 0.255$, $p = 0.003$) and SI ($\beta = 0.274$, $p = 0.015$), but not attitude. Explanatory power was assessed using $R^2$ values for each endogenous construct. Attitude ($R^2 = 0.817$) and BI ($R^2 = 0.789$) demonstrated high explanatory power, while PBC ($R^2 = 0.642$) showed moderate-to-high explanatory power and SI ($R^2 = 0.213$) showed low-to-moderate explanatory power. Figure 2 presents the estimated structural model with standardized path coefficients and $R^2$ values.

One unexpected finding was the positive effect of compatibility on SI. Hypothesis H12 predicted a negative relationship, reasoning that greater compatibility would reduce reliance on other pedestrians' cues. Instead, the positive effect suggests that higher perceived compatibility may increase the extent to which individuals incorporate social information into their crossing decisions. Table 5 summarizes all direct effects and hypothesis outcomes.

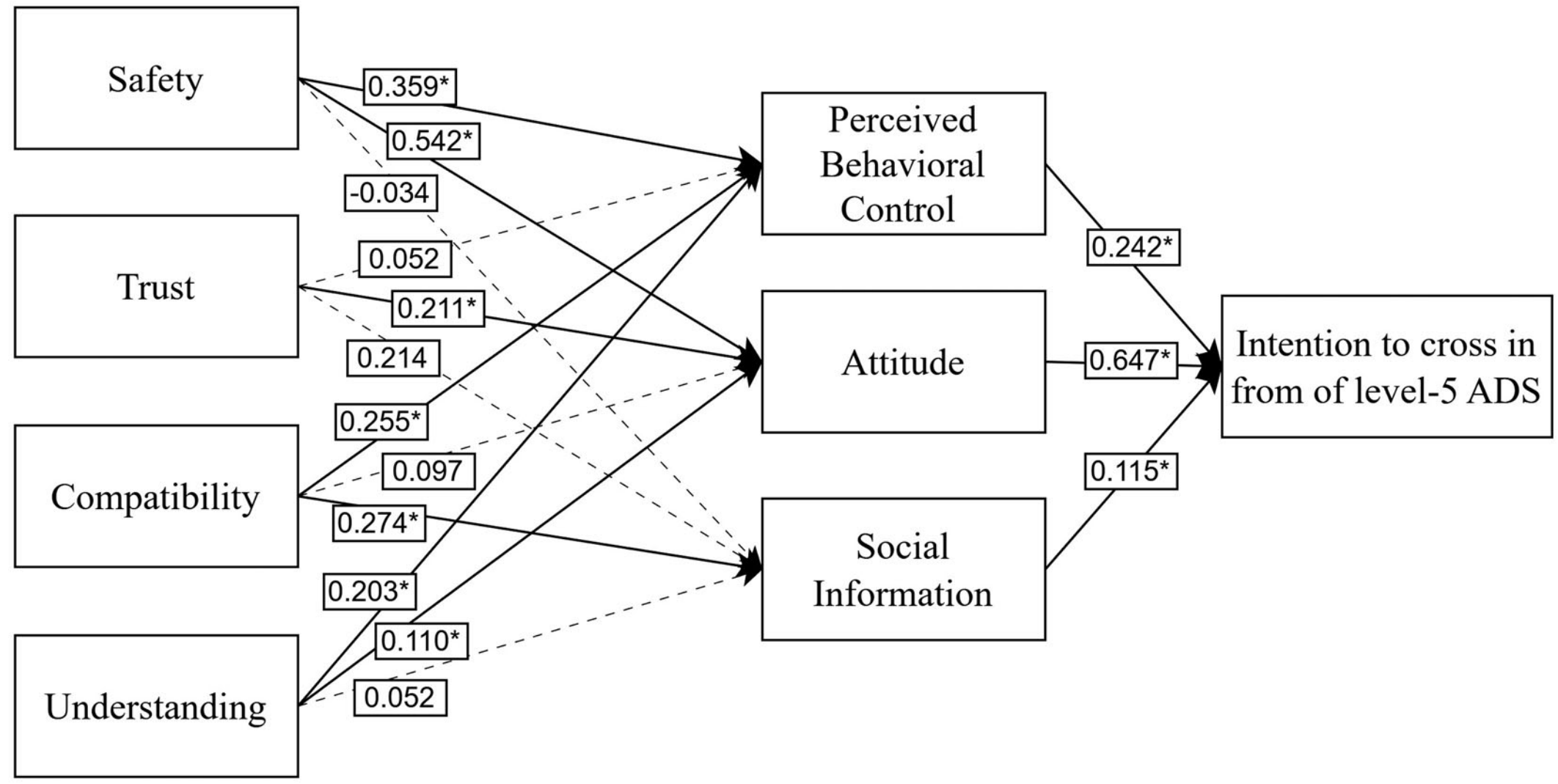

**Figure 2.** Assessment of the structural model. Dashed lines indicate a non-significant path. Standardized path coefficients. $^*p < 0.05$.

**Table 5.** Summary of direct effects, including *t*-values, *p*-values, and effect sizes for the relationships in the structural model.

| Path | Direct effects | *t* | *p* | Hypothesis | Supported? |
|---|---|---|---|---|---|
| PBC → BI | 0.242 | 5.184 | 0.000 | H1 | Yes |
| Attitude → BI | 0.647 | 13.324 | 0.000 | H2 | Yes |
| SI → BI | 0.115 | 2.748 | 0.006 | H3 | Yes |
| Safety → PBC | 0.359 | 3.437 | 0.001 | H4 | Yes |
| Safety → Attitude | 0.542 | 7.990 | 0.000 | H5 | Yes |
| Safety → SI | −0.034 | 0.160 | 0.873 | H6 | No |
| Trust → PBC | 0.052 | 0.494 | 0.622 | H7 | No |
| Trust → Attitude | 0.211 | 3.186 | 0.001 | H8 | Yes |
| Trust → SI | 0.214 | 1.896 | 0.058 | H9 | No |
| Compatibility → PBC | 0.255 | 2.973 | 0.003 | H10 | Yes |
| Compatibility → Attitude | 0.097 | 1.734 | 0.083 | H11 | No |
| Compatibility → SI | 0.274 | 2.431 | 0.015 | H12 | No |
| Understanding → PBC | 0.203 | 2.285 | 0.022 | H13 | Yes |
| Understanding → Attitude | 0.110 | 1.988 | 0.047 | H14 | Yes |
| Understanding → SI | 0.052 | 0.379 | 0.705 | H15 | No |

Effect sizes ($f^2$) were interpreted according to Cohen's (1988) guidelines, where values of 0.02, 0.15, and 0.35 represent small, medium, and large effects, respectively. As shown in Table 6, the effect of attitude on BI was large ($f^2 = 0.970$), and the effect of safety on attitude was moderate-to-large ($f^2 = 0.317$). PBC showed a small-to-medium effect on BI ($f^2 = 0.140$). Several additional predictors demonstrated small effects, including compatibility on PBC ($f^2 = 0.059$), SI on BI ($f^2 = 0.051$), trust on attitude ($f^2 = 0.061$), and understanding on PBC ($f^2 = 0.033$). The remaining effects were below the 0.02 threshold and therefore considered negligible.

Model fit was evaluated using the standardized root mean square residual (SRMR). The saturated model yielded an SRMR of 0.092 and the estimated model an SRMR of 0.093, both below the suggested threshold of 0.10, indicating acceptable model fit (Hair, Hult, et al., 2023).

## 6. Discussion

This study investigated how the Theory of Planned Behavior constructs and external factors (safety, trust, compatibility, and understanding) impact pedestrians' intentions to cross the road in front of a level-5 ADS in an extended TPB model.

**Table 6.** Effect size ($f^2$) estimates for structural model paths.

| Predictor | $f^2$ (Original sample) | $p$-Value |
|---|---|---|
| PBC → BI | 0.140 | 0.016 |
| AT → BI | 0.970 | 0.000 |
| SI → BI | 0.051 | 0.180 |
| S → PBC | 0.071 | 0.136 |
| S → AT | 0.317 | 0.001 |
| S → SI | 0.000 | 0.986 |
| T → PBC | 0.002 | 0.898 |
| T → AT | 0.061 | 0.134 |
| T → SI | 0.014 | 0.393 |
| C → PBC | 0.059 | 0.175 |
| C → AT | 0.017 | 0.456 |
| C → SI | 0.031 | 0.255 |
| U → PBC | 0.033 | 0.321 |
| U → AT | 0.020 | 0.349 |
| U → SI | 0.001 | 0.936 |

### 6.1. Theoretical implications

The model demonstrated strong predictive power while revealing how external perceptions influence the core TPB constructs rather than directly driving behavioral intention. All three TPB constructs (PBC, attitude, and SI) significantly influenced BI. Among the external factors, safety primarily influenced intention indirectly through its effects on attitude and PBC, with trust, compatibility, and understanding contributing additional, generally smaller, effects. Trust was associated with more positive attitudes, compatibility with greater PBC and attention to SI and understanding with modest but significant relationships with both attitude and PBC. These patterns are broadly consistent with findings from Kaye et al. (2022), Evans and Norman (1998), Zhao et al. (2022), and Marcus et al. (2024), and highlight the joint importance of these factors in Level-5 ADS–pedestrian interactions.

The significant impact of social information on behavioral intention aligns with previous studies that found an influence from social information on pedestrians' road-crossing decisions (Faria et al., 2010; Pfeffer & Hunter, 2013; Rosenbloom, 2009). This finding refines the conclusions of Pillai (2017), Jayaraman et al. (2020), and Zhao et al. (2024), who observed that pedestrians commonly expect Level-5 ADS vehicles to behave like considerate human drivers by consistently yielding the right of way. Our findings suggest that even with this expectation, pedestrians still incorporate social cues into their crossing decisions, still indicating relevance in social factors in pedestrian behavior. A study by Zhao et al. (2024) found that pedestrians were less reliant on social information when interacting with automated vehicles if they were equipped with an external human-machine interface (eHMI), a display made up of visual and/or audio features meant to communicate the ADSs' intent to other road users; this suggests that the presence of additional confirming information about the vehicle's intent reduced reliance on social information. Although the study did not incorporate an eHMI, it nevertheless revealed a comparable trend: higher levels of perceived trust and safety were associated with a reduced reliance on social cues within the model. These results suggest that future research ought to further examine the role of social information in pedestrian interactions with level-5 ADSs.

### 6.2. Practical implications

The results offer concrete guidance for the human-centered design of Level-5 ADS vehicles. Perceived safety played a central role in pedestrians' crossing decisions, highlighting the need for Level-5 ADSs to make their intentions explicit, legible, and easy to interpret. Rather than simply behaving safely in the background, these vehicles must *communicate* their awareness of pedestrians and clearly indicate what they are about to do next—whether yielding, waiting, or preparing to proceed. This aligns with prior findings identifying safety communication as a primary concern for vulnerable road users (Deb et al., 2017). In our survey, participants reported moderate agreement with feeling safe crossing in front of a level-5 ADS (S1, $M = 3.05$), but were less confident when asked if level-5 ADSs would pose minimal risk to road users (S3, $M = 2.90$) or if crossing in front of them would involve minimal risk (S4, $M = 2.79$). Particularly telling was the slightly lower agreement when comparing safety with human-driven vehicles (S2, $M = 2.85$), suggesting that pedestrians remain somewhat skeptical about the comparative safety advantages of automated systems. These results show that Level-5 ADSs must not only operate safely but should also communicate their intentions in a clear and accessible manner to maximize acceptance. External human–machine interfaces have been explored for this purpose, and features such as exterior speed indicators, pedestrian detection confirmations, text-based intent displays, and other visual cues can help pedestrians understand when the vehicle has seen them and what it plans to do next.

Compatibility also played an important role, significantly affecting pedestrians PBC and, increasing, rather than decreasing, their attention to social information. This suggests that pedestrians may continue to rely on the behavior of nearby people even when ADSs function smoothly within traffic. Designers should therefore focus on features that help Level-5 ADSs appear consistent with existing traffic norms. This could include maintaining recognizable yielding distances, aligning ADS movement with standard crosswalk rules, and ensuring that eHMI cues do not conflict with pedestrian signal lights. Survey responses indicated general agreement that Level-5 ADSs fit within current infrastructure (C2, M = 3.06), though there was slight disagreement when asked if they thought level-5 ADSs would be compatible with all aspects of the transportation system (C4, M = 2.62). Addressing these concerns may require clearer integration of ADS behavior with intersection logic, lane markings, and pedestrian expectations of local traffic norms.

Trust was also a significant factor, exerting a meaningful influence on attitude, which aligns with findings from Deb et al. (2017) and Jayaraman et al. (2020). However, contrary to our hypotheses, trust did not significantly predict PBC or SI. This pattern suggests that trust shapes general impressions of Level-5 ADSs but does not necessarily guide moment-to-moment decisions in crossing situations. Pedestrians moderately trusted the general reliability of level-5 ADSs (T1, M = 3.10), with overall trust in these systems remaining modest (T4, M = 2.91), and comfort levels for engaging in distracted behaviors while crossing low (T3, M = 1.89). These results indicate that building trust will require consistent, transparent behavior that communicates reliability, while also recognizing that pedestrians are likely to remain attentive and cautious regardless of how much they trust the system. Practical steps toward achieving this include the use of eHMI elements described above, such as visual messages that clarify ADS intent and detection acknowledgments that reassure pedestrians that they have been seen. These types of signals can strengthen trust by making ADS behavior more predictable and easier to interpret. Trust can also be reinforced when automated vehicles behave in ways that align with familiar traffic norms, as mentioned previously, since consistency with existing roadway expectations helps pedestrians judge the system as reliable and competent.

When pedestrians are able to anticipate and interpret a vehicle's intentions, they generally develop more favorable attitudes toward the vehicle and experience a heightened sense of control during the interaction. When pedestrians can predict and comprehend a vehicle's intentions, they develop more positive attitudes and feel greater agency in their interactions. Other studies including eHMIs have found that these displays can support pedestrians' perceptions of safety and trust when interacting with level-5 ADSs, though more research needs to be done to understand what eHMI features are most supportive of efficient interactions (Zhao et al., 2024).

The results of this study suggest that level-5 ADSs need to follow user-centered design principles, with a primary focus on safety features and behaviors that reassure pedestrians. Given the moderate safety ratings across survey responses, manufacturers should prioritize both actual safety mechanisms and their perceptibility to pedestrians. Additionally, these systems should align with existing traffic infrastructure and pedestrian expectations, while providing clear communication of their intentions. Instead of aiming to strip away social dynamics, designers of Level-5 ADSs should plan for how the technology will operate within pedestrians established social landscape—acknowledging that people will continue to depend on social cues even as automation advances.

### 6.3. Limitations and future work

While this study contributes to the understanding of pedestrian interaction with automated vehicles, it faced several limitations that can be addressed through future work. The study sample was recruited through convenience sampling and was primarily US-based, limiting sample diversity and generalizability, especially to other countries, as previous studies (e.g., Lanzer et al., 2020) have shown that differing traffic laws and cultural norms greatly impact pedestrian behaviors. As the survey did not collect participants' country of residence or origin, we cannot quantify how many respondents were located in the United States or abroad; we note the US-based nature of the sample solely due to recruitment through US academic and professional networks and acknowledge the absence of geographic data as an additional limitation. Although we gathered demographic details such as age, gender, disability status, and prior experience, the modest sample of 212 participants restricted our capacity to conduct more fine-grained subgroup analyses, and future studies should investigate how these factors influence decision making around level-5 ADSs, particularly disability status, which remains underexplored.

This study also highlights the need for experimental research into the role of eHMIs in supporting clear, effective communication between pedestrians and level-5 ADSs. Previous studies have found that the inclusion of eHMIs can support perceptions of factors such as attitude, safety, trust, and understanding (Marcus et al., 2025; Zhao et al., 2024). Similarly, the role of SI should be further explored, given the mixed impact seen in the literature (Faria et al., 2010; Pfeffer & Hunter, 2013; Rosenbloom, 2009; Zhao et al., 2024). The unexpected finding that higher perceived compatibility is associated with increased use of social cues warrants deeper investigation, and future observational research could clarify how pedestrians weigh technological signals against social information when deciding whether to cross.

Additionally, this study focused on pedestrians' cognitive and perceptual factors through self-reported intention rather than directly modeling implicit vehicle–pedestrian communication such as approach speed, yielding style, or temporal gaps. Future work should incorporate controlled vignette-based or video-based HCI experiments to evaluate how such kinematic cues interact with the psychological factors identified here. This would allow researchers to connect cognitive predictors with real-time behavioral responses and observed crossing choices, offering a more complete understanding of pedestrian–ADS interaction.

## 7. Conclusion

This study developed and validated a behavioral model that explains how safety, trust, compatibility, and understanding influence pedestrians' road-crossing intentions by operating through attitude, perceived behavioral control, and social information. PBC, attitude, and social information, the three TPB constructs, all showed significant effects on intention. Safety demonstrated a significant influence on both attitude and perceived behavioral control. Trust significantly influenced attitude, while compatibility had a significant effect on perceived behavioral control and, unexpectedly, increased reliance on social information. Understanding also demonstrated significant, though comparatively weaker, relationships with both attitude and perceived behavioral control.

These findings illustrate how technological and social factors jointly influence pedestrians' crossing decisions. They also emphasize the importance of designing level-5 ADSs with clear safety features that reassure pedestrians, ensuring compatibility with existing traffic infrastructure and pedestrian habits, and providing effective communication of vehicle intentions. Such design considerations can help pedestrians feel more confident and in control during interactions with automated vehicles. By accounting for these influences, designers and manufacturers can engineer level-5 ADSs that better align with vulnerable road users' needs and expectations, helping to curb pedestrian–vehicle conflicts and promote safer streets for all.

## Acknowledgment

No funding was received for conduction this study.

## Author contributions

CRediT: **Sanaz Motamedi**: Conceptualization, Investigation, Methodology, Project administration, Resources, Supervision, Validation, Writing – original draft, Writing – review & editing; **Viktoria Marcus**: Conceptualization, Formal analysis, Investigation, Methodology, Visualization, Writing – original draft; **Griffin Pitts**: Formal analysis, Methodology, Visualization, Writing – review & editing.

## Disclosure statement

No potential conflict of interest was reported by the author(s).

## About the authors

**Sanaz Motamedi** is an assistant research professor at Pennsylvania State University, specializing in human-systems engineering and human-automation interaction. Before joining Penn State, she spent five years at the University of Florida teaching and conducting research, and two years as a postdoctoral scholar at the University of California, Berkeley.

**Viktoria Marcus** is a student in the Department of Industrial and Systems Engineering at the University of Florida, with a concentration in human systems. Her research interests include human-technology interaction and technology acceptance in the contexts of automation, online learning, and artificial intelligence.

**Griffin Pitts** is a graduate student in the Department of Computer Science at North Carolina State University. He recently earned a Bachelor of Science in Computer Science from the University of Florida in 2025, where he conducted research on how people interact with, adopt, and trust emerging technologies.